\documentclass{elsarticle}

\usepackage{graphicx}
\usepackage{mathptmx}
\usepackage{algorithm}
\usepackage[noend]{algpseudocode}
\usepackage{graphicx}
\usepackage{tabularx}
\usepackage{url}
\usepackage{amsmath}
\usepackage{amsthm}
\usepackage{amssymb}
\usepackage{subfig}

\usepackage{color}

\newcommand{\ORFEL}{{\bf $ORFEL$}}

\newcommand{\recomms}{Recomms}
\newcommand{\recommendation}{recommendation\ }
\newcommand{\recommendations}{recommendations\ }

\DeclareMathOperator*{\argmax}{arg\,max}

\usepackage{lineno,hyperref}
\modulolinenumbers[5]

\theoremstyle{definition}
\newtheorem{definition}{Definition}

\journal{Information Sciences}

\begin{document}

\begin{frontmatter}

\title{\Large{\ORFEL}: efficient detection of defamation or illegitimate promotion in online recommendation}

\author{Gabriel Gimenes,\\ Robson L F Cordeiro\\ and Jose F Rodrigues-Jr\\}
\address{University of Sao Paulo \\
Av Trab Sao-carlense, 400 \\
Sao Carlos, SP, Brazil-13566-590\\
\{ggimenes, robson, junio\}@icmc.usp.br}

\begin{abstract}
What if a successful company starts to receive a torrent of low-valued (one or two stars) recommendations in its mobile apps from multiple users within a short (say one month) period of time? Is it legitimate evidence that the apps have lost in quality, or an intentional plan (via lockstep behavior) to steal market share through defamation? In the case of a systematic attack to one's reputation, it might not be possible to manually discern between legitimate and fraudulent interaction within the huge universe of possibilities of user-product recommendation. Previous works have focused on this issue, but none of them took into account the context, modeling, and scale that we consider in this paper. Here, we propose the novel method Online-Recommendation Fraud ExcLuder (\ORFEL) to detect defamation and/or illegitimate promotion of online products by using vertex-centric asynchronous parallel processing of bipartite (users-products) graphs. With an innovative algorithm, our results demonstrate both efficacy and efficiency -- over $95\%$ of potential attacks were detected, and \ORFEL\ was at least two orders of magnitude faster than the state-of-the-art. Over a novel methodology, our main contributions are: (1) a new algorithmic solution; (2) one scalable approach; and (3) a novel context and modeling of the problem, which now addresses both defamation and illegitimate promotion. Our work deals with relevant issues of the Web 2.0, potentially augmenting the credibility of online recommendation to prevent losses to both customers and vendors.
\end{abstract}

\begin{keyword}
\texttt{graphs \sep fraud detection \sep defamation \sep recommendation \sep Web 2.0 \sep data analysis}
\end{keyword}

\end{frontmatter}

\section{Introduction}
\label{sec:intro}
In the Web 2.0, it is up to the users to provide content, like photos, text, recommendations and many other types of user-generated information. The more interaction, e.g., likes, recommendation, comments, etc., a product page (or a user profile) gets, the better are the potential profits that a company (or an individual) may achieve with automatic recommendation, advertisement, and/or priority in automatic search engines. In Google Play, for example, mobile apps heavily depend on high-valued (4 or 5 stars) recommendations to get more important and to expand their pool of customers; on Amazon, users are offered the most recommended products, that is, those that were better rated; and in TripAdvisor, users rely on other's feedback to pick their next travels. The same holds for defamation, which is the act of lowering the rank of a product by creating artificial, low-valued recommendations. Sadly, fraudulent interaction has come up in the Web 2.0 -- fraudulent likes, recommendations, and evaluations define artificial interests that may illegitimately induce the importance of online competitors.

Attackers create illegitimate interaction by means of fake users, malware credential stealing, Web robots, and/or social engineering. The identification of such behavior has great importance to companies, not only because of the potential losses due to fraud, but also because their customers tend to consider the reliability of a given website as an indicator of trustfulness and quality. According to Facebook \cite{facebook1}, fraudulent interaction is harmful to all users and to the Internet as a whole, so it is important that users have a true engagement around brands and content.

However, catching up with such attacks is a challenging task, especially when there are millions of users and millions of products being evaluated in a system that deals with billions of interactions per day.
In such attacks, multiple fake users interact with multiple products at random moments \cite{akoglu2008rtm} in a way that their behavior is camouflaged among millions of legitimate interactions per second.
The core of the problem is: {\it how to track the temporal evolution of fraudulent user-product activity since the number of possible interactions is factorial?}

We want to identify the so-called {\it lockstep behavior}, i.e., groups of users acting together, generally interacting with the same products at around the same time. As an example, imagine that an attacker creates a set of fake users to artificially promote his e-commerce website; then, he would like to comment and/or recommend his own Web pages, posts, or advertisements to gain publicity that, fairly, should come from real customers. Here, an attacker may refer to employees related to a given company, professionals (spammers) hired for this specific kind of job, Web robots, or even anonymous users. The weak point in all these possibilities is that the attacker must substantially interact with the attacked system within limited time windows; also, the attacker must optimize his efforts by using each fake user account to interact with multiple products. This behavior agrees with the lockstep definition. Note that this pattern is well-defined in online recommendation and in many other domains, such as academic co-citation, social network interaction, and search-engine optimization.
Provided that this is not a new problem, we use in this paper the definition of {\it lockstep behavior} given by Beutel {\it et al.} \cite{beutel2013copycatch}. See the upcoming Section \ref{sec:problem} for details.

The task of identifying locksteps is commonly modeled as a graph problem -- nodes are either users or products; weighted edges represent recommendations -- in which we want to detect near-bipartite cores considering a given time constraint.
The bipartite cores correspond to groups of users that interacted with groups of products within limited time intervals.
One lockstep may be defamation, when the interactions are negative (low-valued) recommendations; or illegitimate promotion, when the recommendations are positive.
Therefore, the problem generalizes to finding near-bipartite cores with edges whose weights correspond to the rank of the recommendations.
Note that we want to tackle the problem {\it without} any previous knowledge about suspicious users, products, nor the moments when frauds occurred in the past.

This work extends the state-of-the-art solutions for the problem of  lockstep identification. Our main contributions are threefold:

\begin{enumerate}
    \item{{\bf Novel algorithmic paradigm}: we introduce the first {\it \underline{vertex-centric}} algorithm able to spot {\it lockstep behavior} in Web-scale graphs using asynchronous parallel processing; vertex-centric processing is a promising paradigm that still lacks algorithms specifically tailored to its {\it modus operandi};}
    \item{{\bf Scalability and accuracy}: we tackle the problem for billion-scale graphs in one {\it \underline{single}} commodity machine, achieving efficiency that is comparable to that achieved by state-of-the-art works on large {\it \underline{clusters}} of computers, whilst obtaining the same efficacy;}
    \item{{\bf Generality of scope:} we tackle the problem for real weighted graphs ranging from social networks to e-commerce recommendation, expanding the state-of-the-art of lockstep semantics to discriminate defamation {\it \underline{and}} illegitimate promotion.}
\end{enumerate}

This paper follows a traditional organization.
Section \ref{sec:back} presents background concepts,
while Section \ref{sec:related} reviews the related works.
Our proposal is described in Section \ref{sec:methodology}.
In Section \ref{sec:experiments}, we report experimental results, including real data analyses.
Finally, Section \ref{sec:conclusions} concludes the paper and presents ideas for future work.

\section{Background}
\label{sec:back}

\subsection{Vertex-centric graph processing}
\label{subsec:vertex-centric}
We use in this paper the well-known concept of vertex-centric processing \cite{Malewicz2010}. Given a graph $G=(V,E)$ with vertices labeled from $1$ to $|V|$, we associate a value to each vertice and to each edge -- for a given edge $e=(u,v)$, $u$ is the source and $v$ is the target. With values associated to vertices and edges, vertex-centric processing corresponds to the {\it graph scan} approach depicted in Algorithm \ref{alg:graph-scan}. The values are determined according to the computation that is desired, e.g., Pagerank or belief propagation; we illustrate this fact with hypothetical functions $f$ and $g$ in the algorithm. Evidently, a single scan is not enough for most useful computations, therefore, the graph is commonly scanned many times until a criterion of convergence is satisfied. Graph processing, then, becomes what is defined in Algorithm \ref{alg:graph-processing}.

\begin{algorithm}
\caption{Vertex-centric graph processing}\label{alg:graph-scan}
\hrulefill
\begin{algorithmic}[0]
\Procedure{Graph\_scan(Graph G)}{}
\For{$i = 1\ to\ |V|$}
 \State $set_e \leftarrow$ set of edges adjacent to $V[i]$
 \State $V[i].value \leftarrow f(set_e)$
 \For{each edge e in $set_e$}
    \State e.value $\leftarrow g(V[i].value,e.value)$
 \EndFor
\EndFor
\EndProcedure
\end{algorithmic}
\hrulefill
\end{algorithm}

\begin{algorithm}
\hrulefill
\caption{Graph processing}\label{alg:graph-processing}
\begin{algorithmic}[0]
\Procedure{Graph\_processing}{}
\While{convergence criterion is not satisfied}
 \State Graph\_scan(G)
\EndWhile
\EndProcedure
\end{algorithmic}
\hrulefill
\end{algorithm}

The vertex-centric processing paradigm contrasts with usual graph traversal algorithms, like breadth-first or depth-first searches. While traversal-based algorithms support any kind of graph processing, they are made to work with the entire graph in main memory, otherwise, they would be prohibitively costly due to repeatedly random disk accesses. On the other hand, the vertex-centric processing is limited to problems that can be solved along the direct neighbors of the vertices (or with clever adaptations to such constraint); the good point is that it is well-suited to disk-based processing since it can suitably rely on sequential disk accesses. This kind of processing is not only prone to disk-based processing, but also to parallel processing according to which, each thread can be responsible for a different share of the vertices. This possibility yields to quite effective algorithms.

\subsection{Asynchronous parallel processing}
\label{subsec:parallel}
Many researchers have developed systems to process graphs in large-scale, either using vertex-centric or edge-centric processing; this is the case of systems Pregel \cite{Malewicz2010}, Pegasus \cite{Kang2009}, PowerGraph \cite{gonzalez2012powergraph}, and GraphLab \cite{Low2010}. However, such systems are parallel-distributed, and thus, they demand knowledge, availability, and management of costly clusters of computers. More recently, a novel paradigm emerged in the form of frameworks that rely on asynchronous parallel processing, including systems GraphChi \cite{Kyrola2012}, TurboGraph \cite{Han:2013:TFP:2487575.2487581}, X-Stream \cite{roy2013x} and MMap \cite{lin2014mmap}. Such systems use disk I/O optimizations and the neighborhood information of nodes/edges in order to set up algorithms that can work in asynchronous parallel mode; that is, it is not required that their threads advance synchronously along the graph in order to reach useful computation. This approach has demonstrated success to tackle many problems, such as Pagerank, connected components, shortest path, and belief propagation, to name a few. In this paper, we use vertex-centric graph processing over framework GraphChi; however, our algorithm can be adapted to any of the frameworks available in the literature.

\section{Related works}
\label{sec:related}

\subsection{Clustering}
\label{subsec:clustering}
The identification of lockstep behavior refers to the problem of partitioning both the rows and the columns of a matrix -- known in the literature as co-clustering or bi-clustering. Some authors have worked on similar variations of the bi-clustering problem. For example, Papalexakis and Sidiropoulos used PARAFAC decomposition over the ENRON e-mail corpus \cite{papalexakis2011co}; Dhillon {\it et al.} used information theory over word-document matrices \cite{dhillon2003information}, and Banerjee {\it et al.} used Bregman divergence for predicting missing values and for compression \cite{banerjee2007generalized}. Other applications include gene-microarray analysis, intrusion detection \cite{papalexakis2012network}, natural language processing \cite{tu2008unsupervised}, collaborative filtering \cite{Kyrola2012}, and image \cite{cruz2011visual}, speech and video analysis \cite{goyal2010feature}. Note that, in this problem setting, whenever time is considered in the form of time windows to be detected, we have a Non-deterministic Polynomial-time (NP-hard) problem \cite{anagnostopoulos2008approximation} that prevents the identification of the best solution even for small datasets. In fact, we deal with issues fundamentally different from the problems proposed so far, which, according to Kriegel {\it et al.} \cite{kriegel2009clustering}, are not straightly comparable due to their specificities.

Theoretically, our work resembles the works of Gupta and Ghosh \cite{gupta2005robust} and of Crammer and Chechik \cite{crammer2004needle}; similarly, we use local clustering principles, but, differently, we are not dealing with one-class problems. Besides, the core of our technique is a variant of mean-shift clustering \cite{cheng1995mean}, now considering temporal and multi-dimensional aspects. As we mentioned before, our contribution relates not only to performance, but also to a novel algorithmic approach.

\subsection{Detection of suspicious behavior on the Web}
One of the first algorithms tailored to detect suspicious behavior on the Web was designed by Douceur \cite{douceur2002sybil} in 2002. The author coins the term \textit{sybil attack}, in the specific context of peer-to-peer networks. Sybil attacks are attacks in which a single entity can provide multiple identities, that is, a single node in the network can create or steal several other identities and use them to gain advantages, thus, undermining the security of the whole system. Latter, Newsome {\it et al.} \cite{newsome2004sybil} showed that sybil attacks can also occur in sensor networks where the attacker wants to bypass security measures, such as voting mechanisms and resource allocation policies.

One similar type of attack was studied by Chirita {\it et al.} \cite{chirita2005preventing} -- the \textit{shilling attack};
in shilling attacks, fake profiles are used to rate items in a recommendation system.
Chirita {\it et al.} \cite{chirita2005preventing} proposed a technique to analyze profiles and to determine whether or not they are suspicious. Later, Su {\it et al.} \cite{su2005finding} developed an algorithm to detect groups of shilling attacks, in which several profiles act in conjunction to alter the ratings of items in the system.

While both sybil and shilling attacks are similar to the concepts that we propose in this paper, as in defamation and illegitimate promotion, none of the aforementioned works consider the temporal dimension to detect the attacks. Time, in such setting, leads to a different problem with NP-Hard complexity \cite{peeters2003maximum}. Also, none of these works took into account the performance and scale that we consider in this paper.

Other related works use the graph theory to detect suspicious behavior on the Web.
This is the case of algorithm Crochet \cite{pei2005mining} that aims at identifying quasi-cliques based on an innovative heuristic; it is also the case of MultiAspectForensics \cite{maruhashi2011multiaspectforensics} that uses tensor decomposition to detect patterns within communities, including bipartite cores. In another work, Eigenspokes \cite{prakash2010eigenspokes} uses singular-value decomposition to detect unexpected patterns in phone call data; also, Netprobe \cite{pandit2007netprobe} uses belief propagation to find near bipartite cores in e-commerce graphs.
Note, however, that:
in spite of the many qualities of these related works, none of them focuses on performance at the same scale that we do; furthermore, they do not study the same problem that we do here, that is, the detection of a set of users fraudulently interacting with the same set of products at around the same time. In fact, the closest approach to our work is the CopyCatch algorithm \cite{beutel2013copycatch}, which focuses on the unweighted version of the problem in a parallel, distributed setting -- its experimental results reported used one {\it \underline{thousand}} machines. In our work, we introduce a vertex-centric asynchronous parallel algorithm that runs in one {\it \underline{single}} commodity machine, whose performance rivals to that reported in this former work, still achieving similar accuracy rates.

\subsection{Lockstep formulation} \label{sec:problem}
In this section we provide a mathematical description of the lockstep detection problem.
As it was mentioned in the introductory Section \ref{sec:intro}, we generalize the problem by amplifying its scope to defamation and illegitimate promotion.
But before doing so, we present in the following the original formulation for the concept of a lockstep, which was formally defined by Beutel {\it et al.} \cite{beutel2013copycatch} as a temporally-coherent near bipartite core.
Along this section, please refer to Table \ref{tab:symb} for a list of symbols and definitions.

\begin{definition}
A set of products P and a set of users U comprise an {\it $[n,m,\Delta t,\rho]$-temporally-coherent near bipartite core} if and only if there exists $P_i \subset P$ for all $i \in U$ such that:

\begin{align}
|P| &\geq m \\
|U| &\geq n  \\
|P_i| &\geq \rho|P| \ \forall i \in U  \\
(i,j) &\in E \ \forall i \in U, j \in P_i  \\
\exists t_j \in \mathbb{R} \ s.t. \ |t_j - L_{i,j}| &\leq \Delta t \ \forall i \in U , j \in P_i
\end{align}

\label{def}
\end{definition}

\begin{figure}[htb!]
 \centering
    \includegraphics[width=0.7\textwidth]{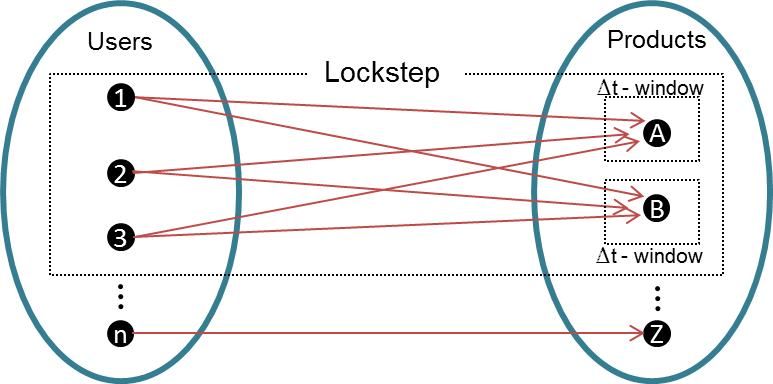}
  \caption{Lockstep illustration: A group of users (1,2 and 3) recommends a group of products (A and B), within limited time-windows for each product, forming a bipartite core.}
     \label{fig:lockstep}
\end{figure}

In other words, we have a lockstep if we find a set of products P that was recommended by a set of users U within a $\Delta t$ time window; we relax this definition with parameter $\rho$, which states that we also have a lockstep if we partially ($\rho$ percentage) satisfy this definition. Note that what makes the problem even more challenging is the temporal factor; also, note that the problem refers to reducing the search space of frauds by pointing out suspicious behaviors, which can turn out to be actually fraudulent, or not. Figure \ref{fig:lockstep} illustrates the concept of a lockstep. It shows how bipartite cores are formed and it also highlights the independence of the time-windows that are exclusive to each product.

While we are considering the aforementioned definition of suspiciousness, it is also important to discuss how effective it is in preventing malicious agents from manipulating recommendations. In other words, we are interested in finding how much damage agents could inflict without being detected. The core fact in the definition of an attack is that: the smaller the attack, the smaller its harm; in consequence, while it is hard to detect very small attacks, they tend to have no use unless they occur in extremely high cardinality. The boundaries of this relation for a non-temporal version of the problem are an open problem, as it is discussed in previous works \cite{beutel2013copycatch,zarankiewicz1951problem}. The challenge becomes even harder when time is considered, as we do in this paper, which sets up an extension of the problem.

In this context, it is possible to conclude that, while we may miss very small attacks and trying to detect them might raise the number of false positives, given the controlled conditions, an adversary can only do a limited amount of damage without getting caught. Finding an optimal strategy of attack is also related to the same open problem discussed before, and since we consider individual temporal centers for each lockstep, one cannot merely use common strategies, such as to wait for a given amount of time before attacking other products with the same users as he/she would be caught anyway. Additionally, the fine tuning of our algorithm's parameters allows the user to detect different types of attacks, further diminishing the possibilities of an adversary to bypassing the system.

\section{Methodology}
\label{sec:methodology}

\subsection{The generalized lockstep problem}
This section shows how to enhance the potential semantics of the lockstep-detection problem by taking into account the weights of edges (e.g., recommendations' scores).

\begin{table}
    \centering
    \scriptsize
        \begin{tabularx}{8cm}{|c | X|}
            \hline
            Symbol & Definition \\
            \hline \hline
            $M$ and $N$ & Number of nodes in each side of the bipartite graph.\\
            \hline
            C & Set of locksteps.\\
            \hline
            I & $M \times N$ adjacency matrix.\\
            \hline
            L & $M \times N$ matrix holding the timestamp of each edge.\\
            \hline
            W & $M \times N$ matrix holding the weight of each edge.\\
            \hline
            $U[c]$ and $P[c]$ & Set of users or products in lockstep c.\\
            \hline
            m and n & Minimum number of products and users in the lockstep to be considered valid.\\
            \hline
            $\Delta t$ & Size of the timespan.\\
            \hline
            $\rho$ & Threshold percentage that the cardinality of the sets of products and users must satisfy to be in a lockstep.\\
            \hline
            nSeed & Number of starting seeds for the algorithm to begin searching for locksteps.\\
            \hline
            $\lambda$ and $\kappa$ & Function and threshold used to define defamation and promotion.\\
            \hline
            $\nu_j$ & Current average time of suspicious recommendations to product j. \\
            \hline
        \end{tabularx}
    \caption{Symbols and Definitions.}
    \label{tab:symb}
\end{table}

Given the formulation presented in Section \ref{sec:problem}, we propose new semantics to the problem by considering weights of edges to define the concepts of defamation -- Equation \ref{eq:def} -- and illegitimate promotion -- Equation \ref{eq:ileg}.
These weights correspond, for instance, to the numeric evaluation (score) given by a user to a product in a recommendation website. Our formulation considers the weights to be positive integers, and we use a threshold $\kappa$ to distinguish between defamation and promotion.

\begin{align}
W_{i,j} &\leq \kappa \ , \ i \in U , j \in P_i \label{eq:def} \\
W_{i,j} &\geq \kappa \ , \ i \in U , j \in P_i \label{eq:ileg}
\end{align}

We consider the problem as an optimization problem, whose objective is to catch as many suspect users as possible, while only growing $P$ until parameter $m$ is satisfied.
Our objective function is in Equation \ref{eq:optm}.
The goal is to find $U[c]$ and $P[c]$ to maximize the number of users and their interactions for a given cluster $c$.

\begin{align}
\argmax_{U[c],P[c]} \sum_{i}{q(L_{i,*}|c,W_{i,*}|c,P[c])}
\label{eq:optm}
\end{align}
\hfill where
\begin{align}
q(u,w,P[c]) &=
\begin{cases} \sigma \ \text{if} \ \sigma = \sum_{j \in P[c]}{I_{i,j} \phi(\nu_j, u_j) \lambda(w_j) \geq \rho m } \\ 0 \ \text{otherwise} \end{cases} \label{eq:q} \\
\phi(t_{\nu},t_u) &=
\begin{cases} 1 \ \text{if} \ |t_{\nu} - t_u| \leq \Delta t \\ 0 \ \text{otherwise} \end{cases} \\
\lambda(g_j) &=
\begin{cases} 1 \ \text{if} \ g_j \geq \kappa \\ 0 \ \text{otherwise} \end{cases} \text{for promotion} \label{eq:promo} \\
\lambda(g_j) &=
\begin{cases} 1 \ \text{if} \ g_j \leq \kappa \\ 0 \ \text{otherwise} \end{cases} \text{for defamation}
\label{eq:defam}
\end{align}

Equations \ref{eq:promo} and \ref{eq:defam} refer to our definitions of illegitimate promotion and defamation, respectively, while Equation \ref{eq:q} shows how we incorporate these weight constraints in the original problem, through the definition of a threshold function $\lambda$. That is, we expand the formulation by including new information relative to the weight of these relationships, as well as incorporate such definitions in the objective function, effectively broadening the scope of the problem and its potential applications.

\subsection{Algorithm ORFEL}
\label{sec:alg}
In order to find locksteps, this section presents the {\bf Online-Recommendation Fraud ExcLuder (ORFEL)}, a novel, iterative algorithm that leverages the idea of vertex-centric processing introduced in Section \ref{subsec:vertex-centric} to expand and improve both the scope (weighted graphs) and the efficiency (scalability on a single computer) of current state-of-the-art approaches. Each iteration of our algorithm executes two functions: $updateProducts$ and $updateUsers$ that, respectively, will add/remove products and users from a lockstep that is being identified. The algorithm iterates until convergence -- that is, until sets $P$ and $U$ stabilize for all the locksteps that were found; we consider that a lockstep is stable when no new product or user enters or leaves the lockstep in one iteration, compared to the previous one.
The full pseudo-code of ORFEL is in Algorithm \ref{alg:gc}.\\

\noindent{\bf Initialization}\\
\noindent{The algorithm relies on seeds to search the data space; the general idea is to have each seed inspecting its surroundings looking for one local maximum. Each seed in the algorithm corresponds to one potential lockstep, which comprises a set of products and a set of users. The initial seeds correspond to minimum locksteps, that is, locksteps with one single product, and a few ($\geq 1$) users each -- the only requirement for the initial set of users is that it cannot be an empty set.}

The initialization step randomly chooses products of the dataset, each one corresponding to one seed.
Then, for each product (seed), ORFEL forms initial locksteps by randomly choosing a constant, small number of users that recommended this product.
This is necessary so that the algorithm has initial elements -- $P[i]$ and $U[i]$ for every ith-lockstep -- scattered throughout the search space.
Later, the initial locksteps will grow iteratively in number of products and users.\\

\noindent{\bf Product update}\\
\noindent{In procedure $updateProducts$ -- see Algorithm \ref{alg:odd} -- we only consider vertices that are $products$, so modifications occur in set $P[i]$ only.
This function is called for every product to test if it fits in one of the locksteps; the test is performed for all locksteps. One product enters a given lockstep if at least $\rho$ percent of the users currently in the lockstep recommended that $product$ within a $\Delta t$ time window. To compute this percentage, the algorithm only considers \recommendations that fit the given weight constraint (represented by the $\lambda$ function),
which characterizes either defamation or illegitimate promotion.

For locksteps with $m$ products, that is, those with the maximum number of products, we test if it is worth to swap one of its products for the candidate one.
This test is similar to the one used to add a product, except that, to be swapped, now the candidate product must contain a superset of the set of \recommendations that the current product has. This is a heuristic approach that leads to an additional coverage of the search space because, as we look for supersets of recommendations, the locksteps tend to increase in size.\\

\noindent{\bf User update}\\
Procedure $updateUsers$ -- see Algorithm \ref{alg:even} -- considers only vertices that are $users$,
so it modifies set $U[i]$ only.
Similarly to what is done in step $updateProducts$, we update each lockstep separately by testing if the current $user$ can be added to it. A candidate $user$ will enter a lockstep if it recommends at least $\rho$ percent of the products in the cluster within a $2\delta t$ time window of each of the products' time centers - that is, the average \recommendation time on that product inside the lockstep - and if it fits the desired weight constraint. If the candidate $user$ fills the requirements, it is added to that lockstep. Note that this step allows $users$ outside the actual $\delta t$ time window to enter the lockstep; this is the mechanism that drives the lockstep towards a better local maximum, whenever there exists one. We propose to use $2\delta t$, following empirical evidence obtained from both our work and the state-of-the-art Beutel's \cite{beutel2013copycatch} approach.} \\

\noindent{\bf End iteration}\\
As it can be seen in Algorithm \ref{alg:gc},
we run procedure $endIteration$ right after step $updateUsers$ is complete.
This additional step is described in Algorithm \ref{alg:endIter}.
For all locksteps, the algorithm sorts the $2 \Delta t$ recommendations by their timestamps and scans them sequentially looking for the subset that maximizes the recommendation criterion (number of recommendations); the target subset must fit a $\Delta t$ window. This is the core mechanism of our algorithm; what it does is to let a $2 \Delta t$ time window to take place at first, then, from the corresponding set of recommendations, it selects a subset that maximizes the target criterion. This mechanism is what makes the seeds ``inspect'' their $2\Delta t$ neighborhoods. If the recommendation set changes, a new iteration will lend new products and users to entering/swapping into the lockstep, until convergence. Once a seed finds a local maximum, it stops evolving and does not change anymore.

Note that some seeds may converge sooner than others,
leading to locksteps smaller than parameters $m$ and $n$.
These seeds are considered ``dead'' (no modifications between iterations), so they are ignored by the algorithm.
They can occur from the second iteration on,
after which the number of locksteps (live seeds) becomes smaller than the initial number of seeds.
The algorithm converges when all seeds are ``dead''.

\begin{algorithm}
\caption{ORFEL Algorithm.}
\label{alg:gc}
\hrulefill
\begin{algorithmic}[0]

\Function{\ORFEL}{\textit{$n,m,\rho,\Delta t,nSeeds$}}

\State Initialize \textit{$ U[nSeeds], P[nSeeds] $} \Comment{Initial Seeding}

\Repeat
    \State $U' = U$
    \State $P' = P$

    \For{each product p in $|V|$}
        \State P = updateProducts(p)
    \EndFor

    \For{each user u in $|V|$}
        \State U = updateUsers(u)
    \EndFor

    \State endIteration()

\Until $U' = U$ and $P' = P$

\State \textbf{return} $[U, P]$

\EndFunction

\end{algorithmic}
\end{algorithm}

\begin{algorithm}
\caption{updateProducts}\label{alg:odd}

\hrulefill
\begin{algorithmic}[0]

\Procedure{updateProducts}{\textit{$vertex$}}
    \For{each Lockstep $c \in C$}
        \State $\recomms \leftarrow U[c].edges \cap vertex.edges$
        \State timeCenter $\leftarrow$ avgtime(\recomms)
        \For{each edge e in $\recomms$}
            \If{$|e.time - timeCenter| > \Delta t$ and $\lambda(e.weight)$}
                \State $\recomms = \recomms - \{e\}$
            \EndIf
        \EndFor

        \If{$|P[c]| < m$}
            \If{$(|\recomms| / |U[c]|) \geq \rho$}
                \State P[c] = $P[c] \cup \{vertex\}$
            \EndIf
        \Else
            \For{each product $p \in P[c]$}
                \If{$p.\recomms \subset \recomms$}
                    \State swap = p
                \EndIf
            \EndFor
            \State P[c] = $(P[c] - \{swap\}) \cup \{vertex\}$
        \EndIf
    \EndFor
\EndProcedure

\end{algorithmic}
\end{algorithm}
\hrulefill

\begin{algorithm}
\caption{updateUsers}\label{alg:even}

\begin{algorithmic}[0]

\Procedure{updateUsers}{\textit{$vertex$}}
    \For{each Lockstep $c \in C$}
        \State $\recomms \leftarrow P[c].edges \cap vertex.edges$
        \For{each edge e in $\recomms$}
            \State pCenter $\leftarrow$ avgtime((u, e.vertex), $u\in U[c]$)
            \If{$|e.time - pCenter| > \Delta t$ and $\lambda(e.weight)$}
                \State $\recomms = \recomms - \{e\}$
            \EndIf
            \EndFor
            \If{$(|\recomms| / |P[c]|) \geq \rho$}
                \State U[c] = $U[c] \cup \{vertex\}$
            \EndIf
    \EndFor
\EndProcedure

\end{algorithmic}
\hrulefill
\end{algorithm}

\begin{algorithm}
\caption{endIteration}\label{alg:endIter}

\begin{algorithmic}[0]

\Procedure{endIteration}{}
    \For{each Cluster $c \in C$}
        \For{each product $p \in P[c]$}
            \State Sort U[c] by the time of the \recomms
            \State Scan sorted U[c] for the $2\Delta t$-subset that maximizes the number of \recomms
            \State Remove the users from U[c] that are not in the subset
        \EndFor
    \EndFor
\EndProcedure

\end{algorithmic}
\hrulefill
\end{algorithm}

\subsection{Discussion about the parameters}
\noindent{As it can be seen in Algorithm \ref{alg:gc}, \ORFEL\ has five parameters: $m$, $n$, $\rho$, $\Delta t$ and $nSeeds$.
The first two, $m$ and $n$, respectively refer to the cardinality of $products$ and $users$ that the algorithm verifies when evaluating suspicious locksteps.
Parameter $\rho$ is the minimum percentage (the tolerance fraction) of products $\rho*m$ for the algorithm to state that a bipartition is, in fact, suspicious.
Although one can freely alter the value of $\rho$, based on empirical evidence, we suggest using no less than 80\%, otherwise, the locksteps might degenerate. Note that we use a single value of $\rho$ for both users and products, because, intuitively, this parameter is expected to be nearly the same for the two entities; nevertheless, algorithmically, one could easily adapt our proposal to use different values at the cost of greater computational complexity. Parameter $\Delta t$ defines the time window within which the interactions (recommendations) should take place. Finally, parameter $nSeeds$ refers to the number of seeds that the algorithm will spread through the search space, each one looking for one suspicious lockstep.}

Parameters $m$ and $n$ define the aspects of the suspicious behaviors that we are looking for.
Increasing (or decreasing) the value of $m$ or $n$ means that we want to find suspicious behaviors involving more (or less) $products$ and/or $users$.
These values define what we call ``$AttackSize$'', i.e., the dimensions of the attacks that we presume to exist.
In practice, parameters $m$ and $n$ filter out attacks that are too small and/or too large, what may be desired depending on the domain.
In the experimental Section \ref{sec:experiments}, we evaluate how distinct configurations of $AttackSize$ impact the efficacy of our algorithm.

Parameter $\rho$ makes the algorithm flexible about different types of attacks, including those in which the $users$ attack only a fraction (percentage) of the expected number of products $m$ in the locksteps.
The value of $\rho$ defines how tolerant we want to be with respect to the very definition of suspiciousness. If we set $\rho$ to $1$, only perfect full locksteps would be considered, in which every user recommended every product of the cluster. On the other hand, if we set $\rho$ too low, such as $\rho=0.5$ for instance, only half of the users would have to recommend each product, possibly leading to incorrect assumptions about the concept of suspiciousness.
In practice, $\rho$ defines that the algorithm should have a tolerance around $m$ and $n$.
Note that parameters $\rho$, $m$ and $n$ depend on the semantics of the problem's domain and ought to be different for each application. It is also true for parameter $\Delta t$, which we describe as follows.

Parameter $\Delta t$ is the time span to be defined by the analyst when searching for attacks.
For instance, let us assume the context of attacks in a social network; in this setting, one could argue that a time span as large as a couple of hours is enough to find ill-intended interactions.
On the other hand, in the context of online reviews,
the time span of one week could be more appropriate.
Note that the same reasoning can also be used to define parameters $m$ and $n$.

Finally, parameter $nSeeds$ controls \ORFEL's potential of discovery;
as we show in the experiments (see Figure \ref{fig:seeds-attacks-caught}),
the minimum number of seeds required to analyze one given dataset follows a linear correlation with the data size.

\subsection{Convergence}
\label{sec:convergence}

Our algorithm finds a set of local maxima for the objective function defined in Equation \ref{eq:optm}. Note that this function is bounded, since the sets of users and products are limited. Therefore, convergence depends solely on the behavior of steps $updateProducts$, $updateUsers$ and $endIteration$.

In step $updateProducts$, the algorithm checks if a given product should be added to any of the current locksteps, deciding to include or to swap that product only if it covers more recommendations than what we have so far. As a result, the objective function may only improve or stay unaltered after this step.
On the other hand, step $updateUsers$ attempts to add suspect users to the existing locksteps by extending the size of the time-window, while step $endIteration$ makes sure that only the largest set of users fitting in the best $\Delta t$ time-window is added to each lockstep.
As a consequence, these last two steps can only improve our objective function, by including more users, or leave it unaltered if no users are added.

These observations lead us to conclude that the locksteps grow asymptotically in our algorithm, eventually reaching a local maximum that prevents changes between two iterations.
Therefore, \ORFEL\ always converges despite the data given as input. Besides this theoretical exercise, the convergence of our algorithm is empirically demonstrated in Section \ref{sec:experiments}.

\subsection{Computational cost}
\label{subsec:compcost}
\noindent{
To study the computational cost of our algorithm,
let us assume that the graph $G$ received as input has size $D$ bytes;
we have $M$ bytes available in main memory, and;
the disk blocks have $b$ bytes each.
In this setting,
\ORFEL\ splits the graph into $\lceil P=D/M \rceil$ parts. Each part contains edges that are sorted in disk according to their source vertices so that the graph is processed by reading the parts twice, first as targets and then as sources.
Therefore, in order to read the entire graph, it is necessary to read $B=D/b$ disk blocks twice, or $2B$ times.
For each part it is also necessary to read the other $P-1$ parts, leading to $P^2$ disk seeks.
Therefore, the cost of disk operations is given by $P^2\ disk\ seeks\ + 2B\ block\ reads$ per iteration}.

\ORFEL\ runs for $I$ iterations.
In each iteration, besides the disk operations, it runs once for each of the $S$ seeds (worst case) processing in memory all the $|E|$ edges of the graph at each time.
Therefore, the processing cost of the algorithm is $I*O(S*|E|)$.

Each iteration of the algorithm asks for a reorganization step in which the locksteps of each seed are redefined based on the results annotated in the last iteration. For $I$ iterations, $S$ seeds, $n$ users and $m$ products, this step runs at cost $I*O(S*n*(m*log(m)))$. Part of this cost is due to the operation of sorting in memory (logarithmic time). This is the worst case scenario, when the algorithm processes all seeds -- the cost drops abruptly after a few iterations because the majority of the seeds does not grow; instead, they stop evolving at a local maximum that is too small to be considered a lockstep, being ignored in further iterations.

Finally, the total cost of \ORFEL\ is $I*(P^2\ disk\ seeks\ + 2B\ block\ reads + O(S*|E|) + O(S*n*m*log(m)))$.
Note that the cost of processing is irrelevant,
since it is $6$ orders of magnitude smaller than that of a mechanical disk and $4$ orders smaller than that of a solid-state disk.
As so, the main cost of \ORFEL\ is $I*(P^2\ disk\ seeks\ + 2B\ block\ reads)$.
Note, from our analysis, that the computational cost depends on the amount of main memory available, which is used as a buffer for data coming from disk;
hence, all the runtime measurements reported in the next section could be smaller if we had more memory to use.

\section{Experiments}
\label{sec:experiments}
\subsection{Experimental setting}
\noindent{We implemented \ORFEL\ using Java 1.7 over the GraphChi platform, as stated in Section \ref{subsec:parallel}. We ran our experiments on an i7-4770 machine with 16 GB of RAM, and 2TB 7200RPM HDD; for the tests with SSD, we used a 240GB drive with I/O at 450MB/s.
For full reproducibility, the complete experimental setup is publicly available at \url{www.icmc.usp.br/pessoas/junio/ORFEL/index.htm}, including source codes and graph/lockstep generators.}

We studied two real-world graphs: $Amazon.FineFoods$ and $Amazon.Movies$.
They are publicly available at the Snap project \cite{snapnets} web page at \url{snap.stanford.edu/}.
Both datasets comprise user-product recommendation data from the Amazon website; the first one refers to the section of fine food products and the second one contains reviews of movies. For each review, we have the corresponding timestamp and one numeric evaluation (score) ranging from 1 to 5.
Synthetic graphs were also studied so to generalize the scope of our tests.
To generate the data, we used a bipartite graph generator that works based on the $G_{nmk}$ model available on NetworkX \cite{hagberg-2008-exploring}, in which $n$ stands for the number of nodes in the first bipartite set; $m$ stands for the number of nodes in the second bipartite set; and $k$ is the number of randomly generated edges connecting both sets.
Table \ref{tab:datasets} lists the two Amazon datasets and the synthetic dataset $Synthetic.C$, which was generated using $n=2,000,000$, $m=8,000,000$ and $k=100,000,000$. Additionally, we generated benchmark datasets that are larger versions of dataset $Synthetic.C$; they were used to study the scalability of our algorithm, as it is described in Section \ref{subsec:efficiency}.\\

\begin{table}
    \centering
    \small
        \begin{tabularx}{10.47cm}{|l | c | c | c | c |}
        \hline
            Dataset & \# Users & \#Products & \# Total nodes & \# Edges\\
            \hline
            Amazon.FineFoods & 256,059 & 74,258 & 330,317 & 568,454\\
            Amazon.Movies & 889,176 & 253,059 & 1,142,235 & 7,911,684\\
            Synthetic.C & 2,000,000 & 8,000,000 & 10,000,000 & 100,000,000\\
            \hline
        \end{tabularx}
    \caption{Datasets.}
    \label{tab:datasets}
\end{table}

\begin{figure*}[h!]
    \centering
    \subfloat[Amazon.FineFoods - Parameters (10,5,0.8,1000) \label{fig:fig1}]{%
      \includegraphics[width=0.5\textwidth]{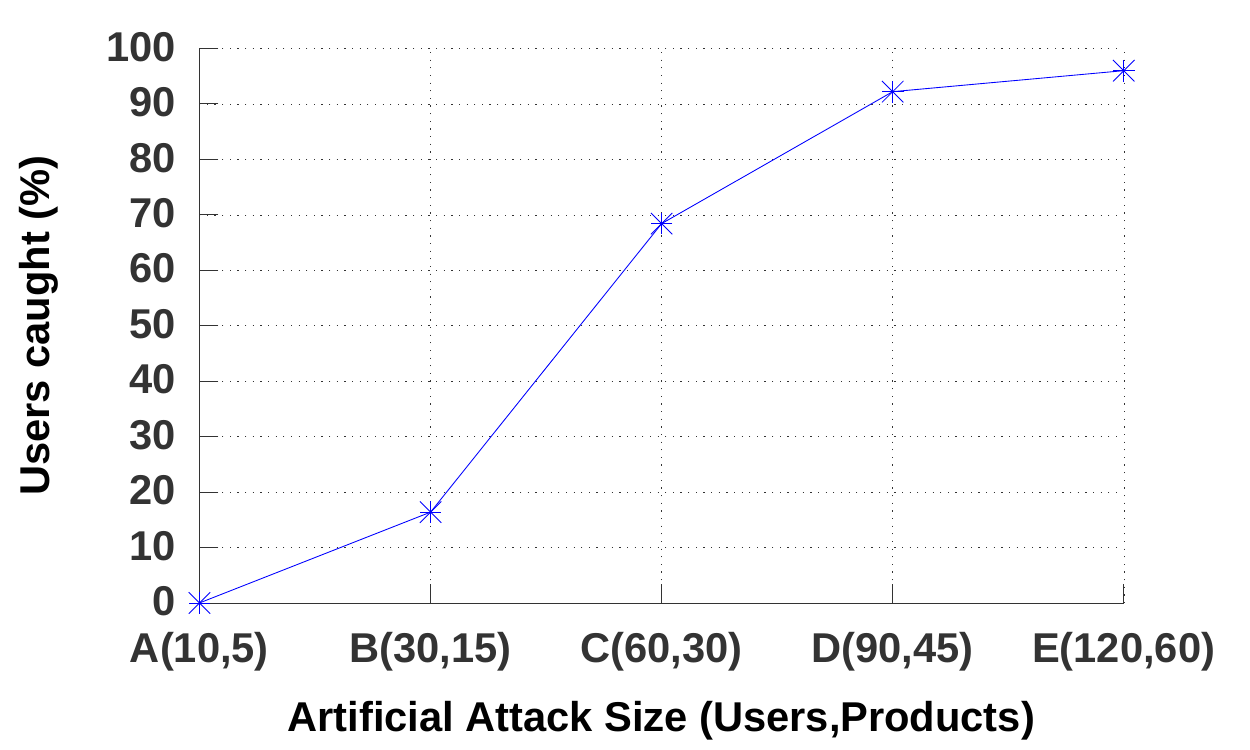}
    }
    \subfloat[Amazon.Movies - Parameters (50,25,0.8,1000) \label{fig:fig2}]{%
      \includegraphics[width=0.5\textwidth]{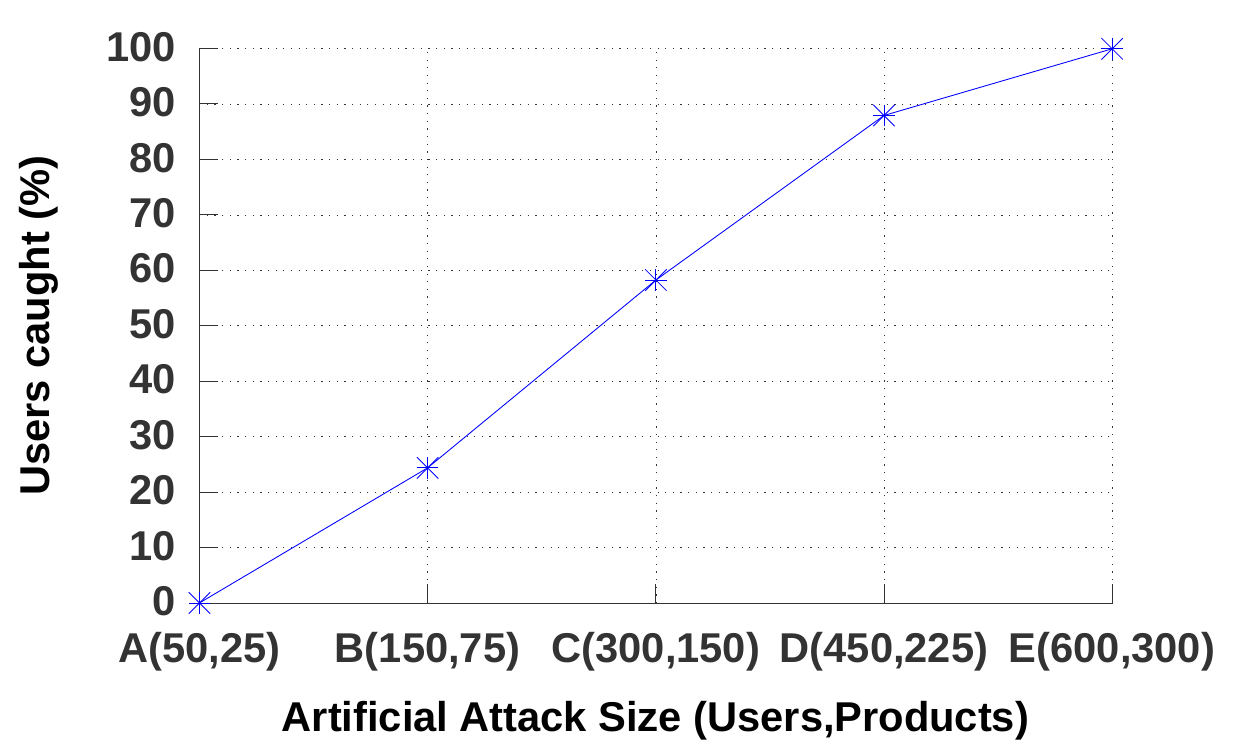}
    }
        \\
    \subfloat[Synthetic.C - Parameters (50,25,0.8,3000) \label{fig:fig3}]{%
      \includegraphics[width=0.5\textwidth]{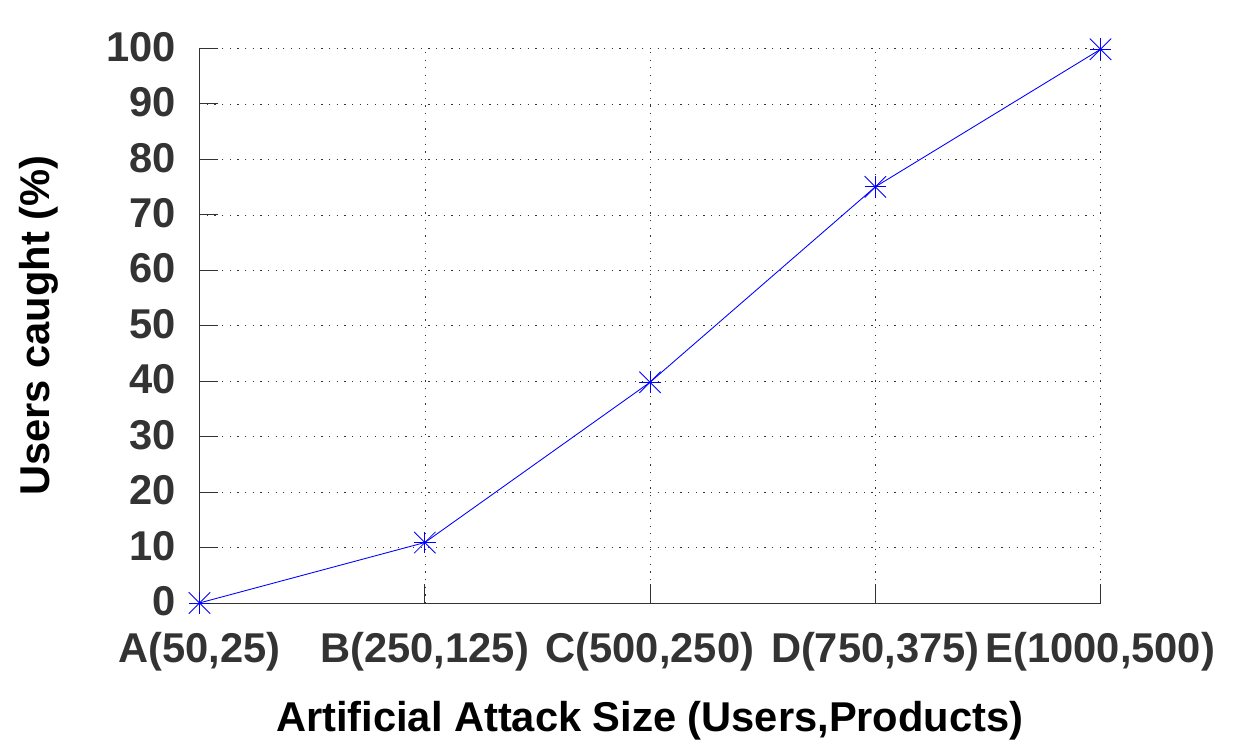}
    }
    \caption{\small{Experiments of efficacy: the percentage of attacks caught versus the size of the artificially generated attacks. Parameters are described as (n,m,$\rho$,nSeeds).}}
    \label{fig:attack-size-attacks-caught}
\end{figure*}

\noindent{\bf Experimental goals}\\
\noindent{The main feature expected from \ORFEL\ is the ability to detect lockstep attacks, either those related to defamation or the ones of illegitimate promotion.
As it was mentioned in Section \ref{subsec:clustering}, this is one NP-hard problem that we approximately solve via an optimization approach. Considering these aspects, we verify: the correctness of our algorithm in Section \ref{sec:initial}; its efficacy (i.e., the ability to find the majority $>95\%$ of the lockstep attacks) in Section \ref{subsec:efficacy}; and; its efficiency (i.e., the ability to reach efficacy within desired time constraints) in Section \ref{subsec:efficiency}.}

\subsection{Preliminar tests under controlled conditions}
\label{sec:initial}
In the first experiment, we used 4 small (thousand-edge scale) synthetic graphs to verify if the algorithm detects locksteps only when they really exist.
These are the controlled conditions of our experimentation, that is, we wanted to make sure that the algorithm would not point out suspect behaviors when we had ascertained that there were none to be detected. We generated synthetic graphs in which no product, nor set of products would configure a given suspect behavior, such as 10 products and 25 users. We then ran \ORFEL\ with varying parameters, including $m=5$, $n=10$ and $m=10$, $n=25$ and verified that, as expected, no lockstep was detected. The same results could be inferred from the algorithm description in Section \ref{sec:alg} and also from the discussion in Section \ref{sec:convergence}, still, we verified this feature empirically.

\subsection{Efficacy}
\label{subsec:efficacy}
\noindent{We define efficacy as the ability to identify the majority ($>95\%$) of the locksteps. To test this feature, we created controlled conditions with artificial attacks appended to our datasets that allowed us to evaluate the output of the algorithm. Algorithm \ref{alg:lockstepper} shows how we generated such attacks, by randomly choosing a group of products and users and then connecting them within a limited $\Delta t$ time window. This was necessary because, since the problem is NP-hard, we would not be able to know whether or not the output of the algorithm is correct considering uncontrolled conditions. This problem is a variation of subspace clustering, considering the semantic that the clusters (locksteps) are unusual and, therefore, suspect. Note that we did not focus on the issue of determining whether or not a given suspect lockstep is actually an attack, since this is one distinct problem that demands extra information (i.e., identification, customer profile, and so on) to be evaluated by means of false-positive and true-positive rates.\\

\begin{algorithm}
\hrulefill
\caption{Lockstep Generator}\label{alg:lockstepper}
\begin{algorithmic}[0]
\Procedure{Lockstepper}{Graph G, nUsers, nPages, $\Delta t$}

\State users = GetRandomUsers(G, nUsers);
\State pages = GetRandomPages(G, nPages);

\For{each Page P $\in$ Pages}
    \State timestamp = getRandomTimeStamp();
    \State rating = getRandomRating();
    \For{each User U $\in$ Users}
        \State newTimeStamp = timestamp + getRandomVariation($\Delta t$);
        \State addEdge(G, U, P, newTimestamp, rating);
    \EndFor
\EndFor
\EndProcedure
\end{algorithmic}
\hrulefill
\end{algorithm}

\noindent{\bf Attack size}\\
\noindent{In order to analyze the ability of the algorithm to find locksteps of different sizes, we ran experiments for each of the three datasets described in Table \ref{tab:datasets}. We fixed $m$ and $n$ in each case while varying the sizes of the artificial attacks appended to the dataset, so to be able to see how effective the algorithm is, depending on the size of the attacks.}

In the first experiment, for each dataset of Table \ref{tab:datasets}, we appended artificial attacks to the data with sizes varying from 10 users and 5 products to 1,000 users and 500 products. This allowed us to observe the percentage of attacks caught for each configuration -- in Figure \ref{fig:attack-size-attacks-caught}(a), ($n=10$,$m=5$,$\rho=0.8$,$nSeeds = 1000$); in Figure \ref{fig:attack-size-attacks-caught}(b), ($n=50$,$m=25$,$\rho=0.8$,$nSeeds = 1000$); and, in Figure \ref{fig:attack-size-attacks-caught}(c), ($n=50$,$m=25$,$\rho=0.8$,$nSeeds = 3000$). One can see a similar behavior in all plots; that is, the percentage of users caught tends to grow as their sizes become larger than the size described by the input parameters given to the algorithm. Intuitively, the larger the attacks are, the more likely that they will be detected using a given configuration. Concomitantly, the smaller the attacks, the less harmful they are. Lastly, notice that even the smaller attacks could be detected with proper parameters -- in this experiment, however, we wanted to demonstrate the general behavior of the algorithm when using specific parameter settings, and not whether  or not smaller attacks could be detected.

This experiment also indicates that ORFEL behaves as expected  for such task in terms of efficacy. That is, if we compare the behavior of ORFEL with that of the state-of-the-art algorithm CopyCatch -- see Figure 6b at \cite{beutel2013copycatch} -- one can see that both approaches present a very similar curve for spotting artificial attacks according to the attack size and the algorithm parameterization.
The results in \cite{beutel2013copycatch} maintain the same intuition regarding which attacks are easier to detect and how important is the parameter tuning,
therefore, they corroborate our conclusions with regard to ORFEL's efficacy.\\

\noindent{\bf Number of seeds}\\
\noindent{
We also used our three datasets to study the behavior of \ORFEL\ regarding the number of seeds that it uses.
We ran each experiment 4 times in each dataset -- as the algorithm is non-deterministic -- and report the average response.
It was a requirement that none of the 4 runs would present discrepant results, and we were able to verify this desirable property since the variance of the results was on the order of 1\%.
We introduced 20 synthetic attacks (10 defamations and 10 illegitimate promotions) in each dataset and varied the number of seeds from 1,000 to 7,000.
Figure \ref{fig:seeds-attacks-caught} reports the results obtained in these experiments.
For dataset Amazon.FineFoods -- the smaller one with 550 K edges -- we were able to catch over $95\%$ of the attacks with 4,000 seeds.
Interestingly, the average of attacks caught with 5,000 seeds was significantly lower, indicating that the algorithm reached its peak performance with nearly 4,000 seeds and only had some variation afterward due to its non-determinism.
Figure \ref{fig:seeds-attacks-caught} also reports that the algorithm caught over $95\%$ of the attacks with 6,000 seeds in dataset Amazon.Movies (8 M edges), and;
it caught over $95\%$ of the attacks with 7,000 seeds in dataset  Synthetic.C (100 M edges).
These results indicate that the best number of seeds
follows a linear correlation with the data size, being approximately $10^3*log({number\ of\ edges})$. For our 3 datasets, it is $\sim5800$, $\sim6900$ and $\sim8000$ respectively.

\begin{figure}[htb!]
 \centering
    \includegraphics[width=0.5\textwidth]{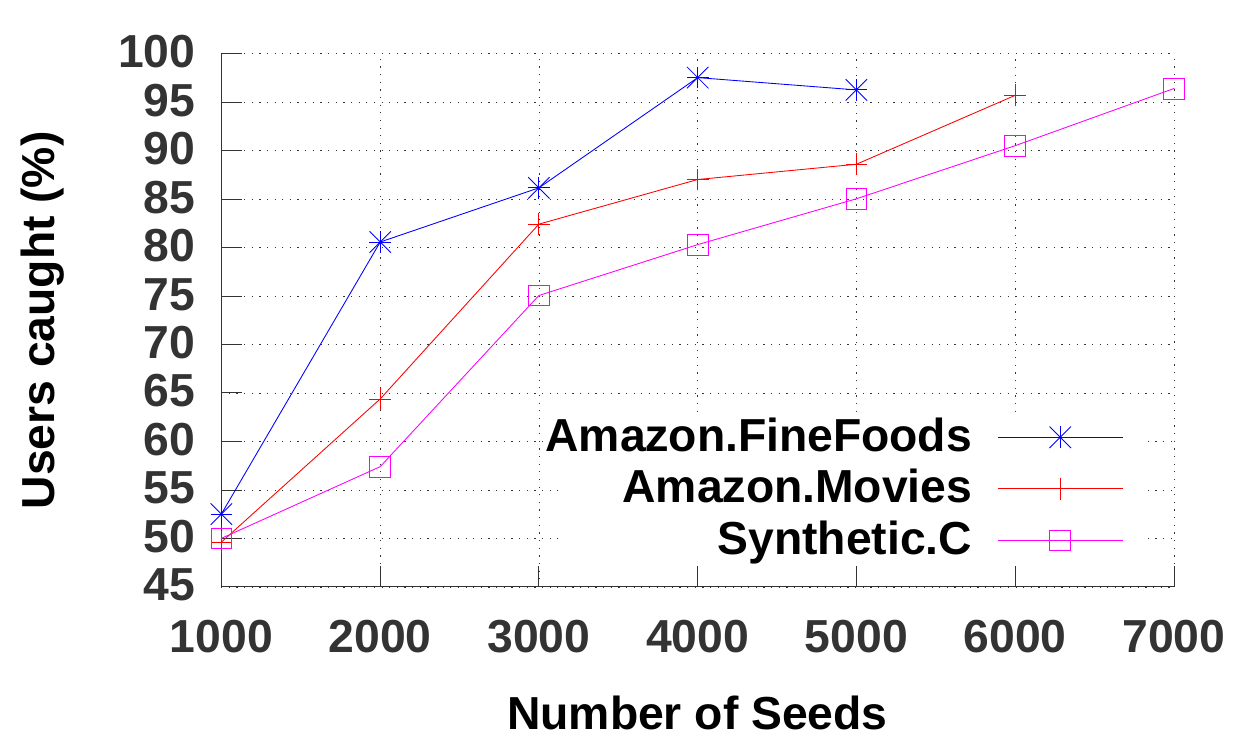}
  \caption{\small{Experiments of efficacy: the percentage of attacks caught versus the number of seeds. Efficacy is demonstrated when over $95\%$ of the attacks are caught. Parameters [n,m,$\rho$,AttackSize(Users,Products)] are: Synthetic.C [50,25,0.8,(750,375)]; Amazon.Movies [50,25,0.8,(500,250)]; Amazon.FineFoods [10,5,0.8,(50,25)]}.}
     \label{fig:seeds-attacks-caught}
\end{figure}

From this experiment we verified that \ORFEL\ is effective; it identified more than $95\%$ of the attacks in three datasets of different sizes. Also, notice that the algorithm accurately detected attacks of distinct sizes, as it can be seen in the parameters of Figure \ref{fig:seeds-attacks-caught}.
It means that ORFEL fits the peculiarities of distinct domains.\\

\noindent{\bf Experiments on real data}\\
\noindent{We performed additional experiments using our real datasets Amazon.Movies and Amazon.FineFoods, this time {\it without including any synthetic data}.
In this context, we considered that a suspect behavior would be 20 users positively recommending 6 movies or food products in less than a week, which, in this semantic context, is an intense load of recommendations. We ran the algorithm and found 37 suspect locksteps; it took 8 minutes to achieve convergence for the largest dataset, Amazon.Movies.
Since the execution time was quite small, we also tested the algorithm using variations of the initial attack description, with 15 users and 7 movies within a week, and also 10 users and 10 movies within three days.
After manually analyzing the suspect locksteps, we discovered that they were caused by amazon's policy of using different identification numbers for different flavors/sizes of the same food product, and for different versions of the same movie, while merging their reviews. Although the locksteps found were not actual attacks, this simple experiment revealed a behavior that should be better analyzed since the aforementioned Amazon's policy could eventually lead customers to misleading choices. In summary, we were able to identify, in a universe of 1,140,000 nodes and 8,000,000 edges, tiny temporal patterns that demand close attention to be noticed. We also emphasize the reduced time required to obtain such results, which allowed us to study different parameter settings according to the data semantics.}

\subsection{Efficiency and scalability}
\label{subsec:efficiency}
\noindent{Due to the current scale of network-like data, our method must be efficient. That is, it must handle billion-scale graphs in reasonable time.
We tested this requirement with synthetic, benchmark datasets that are larger versions of dataset $Synthetic.C$.
Although there are plenty of real data related to our problem (network data, including edge weights and time stamps), such data is rarely shared by companies due to privacy matters.}\\

\noindent{\bf Preprocessing}\\
Asynchronous Parallel Processing platforms like those reviewed in Section \ref{subsec:parallel} demand a preprocessing step in which the data is organized and formatted in accordance to the platform's paradigm. In our case, this step converts text to binary data, then it sorts and writes the vertices in order, so
to have them read from disk with sequential scans, minimizing the number of seeks.
We take nearly 45 minutes, wall-clock time, to preprocess 1 billion edges on a mechanical disk, and nearly 15 min. on a solid-state disk -- for a given dataset, preprocessing is necessary only once, no matter how many times we shall process the data later on.\\

\noindent{\bf Number of edges}\\
\noindent{
We tested the time scalability of our algorithm regarding the number in edges of the input graph.
In the first experiment, we ran \ORFEL\ with 100 seeds;
we took 7 runtime measurements with the number of edges varying from 50 million to 1 billion, each of these measures were obtained as the average of 3 individual runs.
Figure \ref{fig:rutime-edges} reports the results for the mechanical disk; clearly, the runtime scales linearly with regard to the number of edges.
For this configuration, \ORFEL\ took 143 min. ($\approx$2.38 hour) to process 1 billion edges stored on a mechanical disk, and 78 min. ($\approx$1.3 hour) using a solid-state disk.}

We argue that this performance is {\it very efficient} because the previous work (see Figure 4a in Beutel et al. \cite{beutel2013copycatch}) took $\approx$0.5 hour to do a similar processing with {\it \underline{one thousand}} machines over MapReduce, while we used one {\it \underline{single}} commodity machine.
Our gain in performance is considerable
because the former work executes a sequential (non-parallel) algorithm to compute one seed at a time in each machine;
therefore, performance comes at the cost of using thousands of machines, each one executing an instance of the computation, in a distributed environment that has heavy communication demands.
Differently, our algorithm explores the fact that the problem can be solved considering only the neighborhood of each node, thus allowing us to process the graph in a parallel asynchronous mode with multiple seeds being processed simultaneously.
Note that our gains in performance were not only remarkable -- they also made it possible to tackle the problem using commodity hardware.\\

\begin{figure}[htb!]
 \centering
    \includegraphics[width=0.5\textwidth]{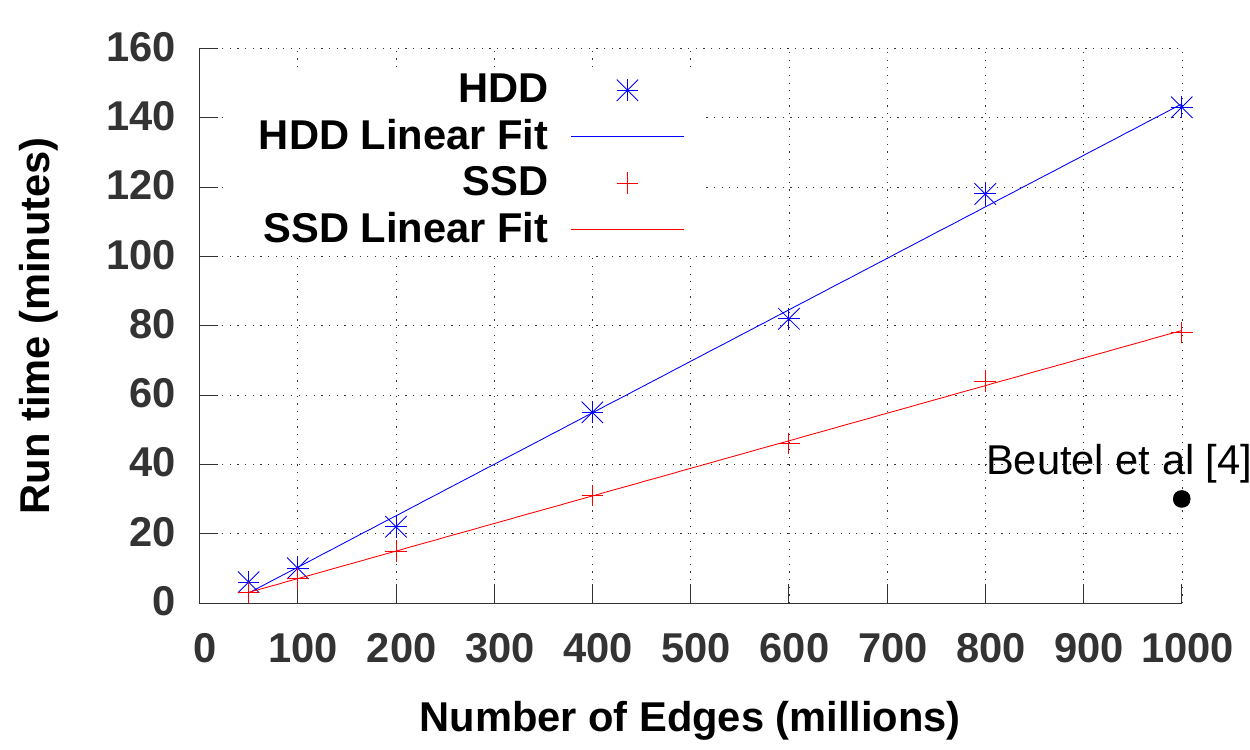}
  \caption{\small{Experiments of scalability on the number of edges: linear growth of computation time (100 seeds) versus the number of edges for mechanical disk (HDD) and for solid-state disk (SSD). Parameters [n,m,$\rho$,AttackSize(Users,Products),nSeeds] are: [50,25,0.8,(500,250),100]}.}
     \label{fig:rutime-edges}
\end{figure}

\noindent{\bf Number of seeds}\\
\noindent{
We also studied the runtime of \ORFEL\ with distinct numbers of seeds.
In this experiment, we used a graph with 100 million edges varying the number of seeds from 100 to 5,000.
Figure \ref{fig:rutime-seeds} reports the results;
as it can be seen, our proposed method scaled linearly in time with regard to the number of seeds.
The algorithm took 10 min. to process the data using 100 seeds, while it took 298 min. with 5,000 seeds;
that is a 30-times increase in runtime for a 50-times increase in the problem input.
}

\begin{figure}[htb!]
 \centering
    \includegraphics[width=0.5\textwidth]{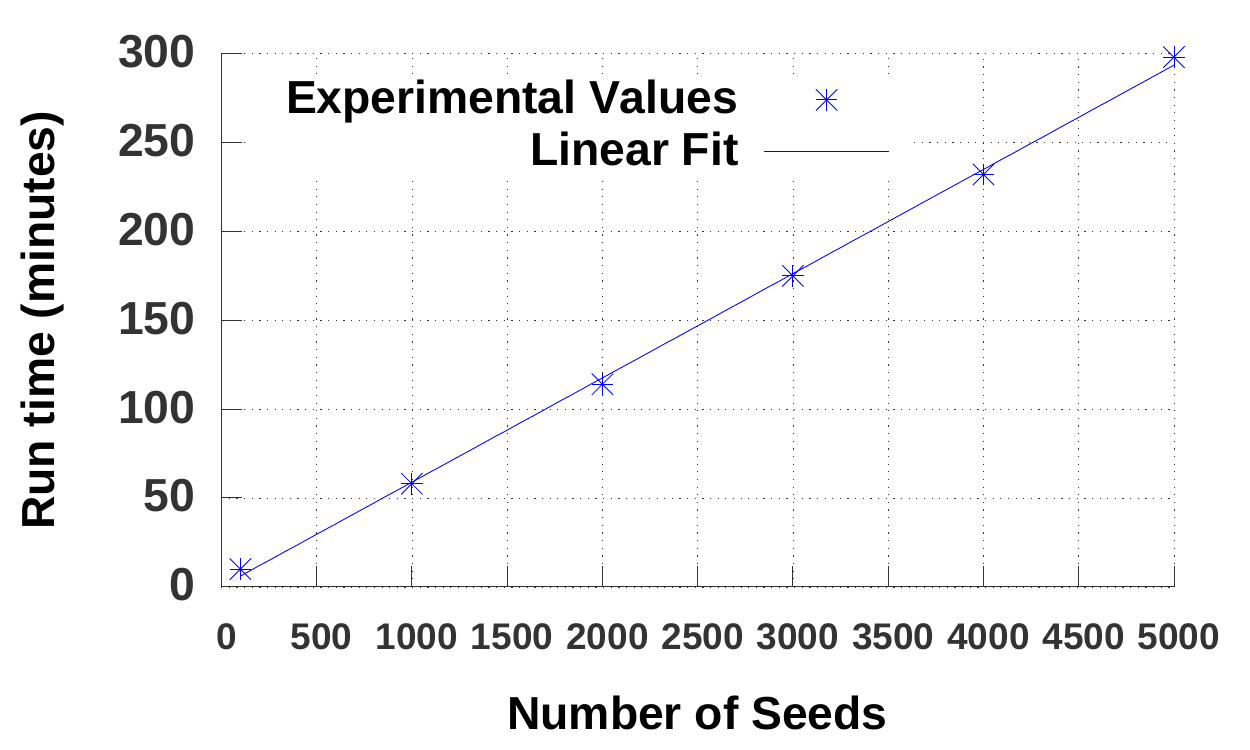}
  \caption{\small{Experiments of scalability on the number of seeds: linear growth (line coefficient $<$ 1) of computation time versus the number of seeds for mechanical disk (HDD) over 100 million edges. Parameters [n,m,$\rho$,AttackSize(Users,Products)] are: [50,25,0.8,(500,250)]}.}
     \label{fig:rutime-seeds}
\end{figure}

\section{Conclusions and future work}
\label{sec:conclusions}

\subsection{Conclusions}
We conclude that, although the problem of detecting fake online interaction over time is NP-hard, it is possible to timely detect most of the malicious activities even with low-cost computer machinery. To do so, we designed a vertex-centric-based graph algorithm using the asynchronous parallel processing paradigm. The general problem is modeled as a bipartite weighted and timestamped graph from which we want to detect temporal near-bipartite cores. This model suits to problems of systematic attacks aimed at defaming or promoting online entities in applications of high-impact, e.g., user-product recommendations, user-app evaluations and journal-journal co-citations, which may be performed by means of fake users, malware credential stealing, Web robots, and/or social engineering.
To validate our proposal, we studied two real graphs of e-commerce, besides synthetic data.

Finally, we emphasize the importance of detecting lockstep behavior, either for defamation or promotion, because these frauds may harm both customers and vendors by inducing sales of unverified products.
We also note that this problem gets even more
relevant as the Web 2.0 expands,
in which the habits of the users are heavily influenced by
online trading and recommendation.

\subsection{Future work}
First, an interesting direction would be further analyzing the boundaries of potential damage an attacker could inflict without being detected, given by the compromise between the size of the attack and the number of attacks. Which as stated in \ref{sec:problem} is an open problem.
Also, as we mentioned before in Section \ref{sec:intro}, \ORFEL\ suits other problems that can be represented as graphs, lending support to additional applications as detailed next.

\subsubsection{Social networks}
\noindent{In social networks, the usual interaction is to like a given post, such as in Google+ or in Facebook;
for this configuration, locksteps characterize solely illegitimate promotion, in which a given post (or page) gets fake likes from attackers willing to make it more relevant than it really is. According to the model of \ORFEL, this problem refers to one unweighted bipartite graph, i.e., all edges weight the same.}

\subsubsection{Journal co-citations}
\noindent{Given the pressure for relevance and impact, some scientific journals may use a co-citation scheme in which one journal cites the other and vice-versa, just like in the case spotted by Nature in 2013 \cite{nature}.
According to this scheme, which is one variant of the lockstep behavior, a journal tends to favor papers that cite a specific journal;
editors may even recommend authors what to cite in their work as a condition for publication.}

To identify this kind of lockstep behavior is not a trivial task because systematic co-citation tends to ``disappear'' along years of publications, provided that such schemes are usually covered by the volume of legitimate citations and by the magnitude of time.
For example, it is reasonable to have co-citation between any two journals in a period of 10 years.
The problem becomes even harder if more than two journals -- e.g., three or four journals -- set up the scheme. In this case, a simple journal-to-journal interaction may not be sufficient to detect the scheme.
The temporal factor and the volume of data make it a problem much harder than simply detecting bipartite subgraphs.

This problem is another instance of the lockstep detection problem studied in our work.
With \ORFEL, it is possible to spot co-citation occurring, let us say, within periods of 1 or 2 years for any number of journals.
For time intervals such as those, one may suspect if a set of journals cite each other with high intensity.

In this specific case, our problem formulation changes a little.
Our model assumes users recommending products -- a bipartite graph;
for detecting journal co-citations we must replicate the set of journals under investigation.
That is, each journal must be represented twice in the model:
once as a citing journal and one other time as a cited journal, thus, defining a bipartite graph as expected by our algorithm.
The output of \ORFEL, then, shall present bipartite subgraphs.
However, distinctly from the user-product model,
it is not enough to identify bipartite subgraphs as an indication of fraud;
we must also have a high similarity between the two sets of nodes in each subgraph.
This similarity can be straightly evaluated using the Jaccard set similarity: $Jaccard=|set_1\cap set_2|/|set_1\cup set_2|$, which returns $1$ if two sets are exactly the same, and $0$ if they have no intersection.
For the co-citation problem, our algorithm could be configured to return the set of bipartite subgraphs ordered by their Jaccard similarity.
Of course, \ORFEL\ spots behaviors that are solely suspicious --  not definitive frauds;
they must go through human interpretation for a definitive decision, considering, for example, that it is expected that the journals with very high impact rates cite each other,
while the same behavior is not expected for journals with lower impact rates.

Note that our algorithm cannot only detect suspicious co-citation cases -- it can do it very efficiently.
Since \ORFEL\ is fast and scalable,
it can virtually inspect {\it \underline{all}} the publication interaction ever produced in just a few hours.

\section{Acknowledgments}
{{\it We thank Prof. Christos Faloutsos and Alex Beutel, from Carnegie Mellon University, for their valuable collaboration.}
This work was funded by Conselho Nacional de Desenvolvimento Cientifico e Tecnologico (CNPq - 444985/2014-0), Fundacao de Amparo a Pesquisa do Estado de Sao Paulo (FAPESP - 2013/10026-7, 2016/02557-0 and 2014/21483-2), and Coordenacao de Aperfeicoamento de Pessoal de Nivel Superior (Capes).}

\bibliographystyle{apa}

\end{document}